\pdfoutput=1
\documentclass[11pt,twoside,a4paper,cmspaper,final,collab]{cms-tdr}
\def\svnVersion{4a3fb49}\def\svnDate{2022/01/27}\def\cmsCernNoTag{CERN-EP-2021-029}\def\cmsCernDate{\today}\def\cmsMessage{Published in the European Physical Journal C as \href{http://dx.doi.org/10.1140/epjc/s10052-022-10095-5}{\doi{10.1140/epjc/s10052-022-10095-5}.}}
\begin{document}\cmsNoteHeader{EXO-17-010}

\newlength\cmsFigWidth
\ifthenelse{\boolean{cms@external}}{\setlength\cmsFigWidth{0.85\columnwidth}}{\setlength\cmsFigWidth{0.4\textwidth}}
\ifthenelse{\boolean{cms@external}}{\providecommand{\cmsLeft}{upper\xspace}}{\providecommand{\cmsLeft}{left\xspace}}
\ifthenelse{\boolean{cms@external}}{\providecommand{\cmsRight}{lower\xspace}}{\providecommand{\cmsRight}{right\xspace}}
\hyphenation{par-ti-cu-lar}

\cmsNoteHeader{EXO-17-010}

\title{Search for strongly interacting massive particles generating trackless jets in proton-proton collisions at \texorpdfstring{$\sqrt{s} = 13\TeV$}{sqrt(s) = 13 TeV}}

\date{\today}

\abstract{{\tolerance=800 A search for dark matter in the form of strongly interacting massive particles (SIMPs) using the CMS detector at the LHC is presented. The SIMPs would be produced in pairs that manifest themselves as pairs of jets without tracks. The energy fraction of jets carried by charged particles is used as a key discriminator to suppress efficiently the large multijet background, and the remaining background is estimated directly from data. The search is performed using proton-proton collision data corresponding to an integrated luminosity of 16.1\fbinv, collected with the CMS detector in 2016. No significant excess of events is observed above the expected background. For the simplified dark matter model under consideration, SIMPs with masses up to 100\GeV are excluded and further sensitivity is explored towards higher masses. \par}}

\hypersetup{
pdfauthor={CMS Collaboration},
pdftitle={Search for strongly interacting massive particles generating trackless jets},
pdfsubject={CMS},
pdfkeywords={CMS, exotics, SIMP, dark matter}}

\maketitle

\section{Introduction}

A major thrust of the experimental programme at the CERN LHC is the search for physics beyond the standard model. In this context, strong emphasis has been placed on the search for dark matter (DM), the nature of which is one of the central questions in particle physics. These DM searches typically target a weakly interacting massive particle (WIMP) with a mass around the electroweak scale. Such a particle can account naturally for the measured DM abundance in the universe, assuming thermal DM production in the $\Lambda$CDM standard cosmological model~\cite{Kolb:1990vq,Bertone:2016nfn}. If produced at the LHC, such a WIMP would, like a neutrino, not be seen in the detector, so would give rise to signatures with transverse momentum (\pt) imbalance.

Because existing searches for WIMPs have excluded much of the parameter space of minimal models, many theoretical developments now extend those models or alter their basic assumptions. In this analysis, we consider the possibility that DM is produced at the LHC, and that its interaction cross section with ordinary matter is so large that the particles are not WIMPs, but rather SIMPs, or strongly interacting massive particles,  whose interactions with nucleons have large cross sections. Such particles could be copiously produced at the LHC, and leave observable signals in the CMS detector. With an interaction cross section as large as the hadronic one, these SIMPs manifest themselves as jets in the calorimeter, but without the presence of tracks from charged hadrons in the tracking detector, in other words as ``trackless jets'', in sharp contrast to typical quantum chromodynamics (QCD) jets. While at first sight it may not seem plausible that such a particle would not have been detected before, it is actually possible to construct a simplified model of SIMPs, interacting through a new scalar or vector low-mass mediator, that evades the many relevant existing bounds~\cite{Daci:2015hca,Bai:2011wy}. In this model, the interaction Lagrangian for a SIMP fermion $\chi$ and a scalar mediator $\phi$ is given by
\begin{linenomath*}
\label{eq:model}
\begin{equation}
\mathcal{L}_\text{int} = - g_\chi \phi \,\overline{\chi} \chi - g_\cPq \phi \,\overline\cPq \cPq.
\end{equation}
\end{linenomath*}
{\tolerance=800 One of the requirements of this model is a purely repulsive SIMP-nucleon interaction with opposite-sign couplings to avoid the formation of bound states between SIMPs and nucleons. At relativistic energies, repulsive and attractive interactions with the same absolute strength have similar behaviour and result in similar kinematics. The coupling strength between the SIMP and nucleons is limited to minimize the impact of the new interaction on the nuclear potential. Furthermore, a scenario with fermionic, asymmetric DM, where no dark antimatter remains, must be considered to avoid excessive Earth heating and neutron star collapse~\cite{Bai:2011wy}. \par}

\begin{figure}[h!]
  \centering
  \includegraphics[width=0.44\textwidth]{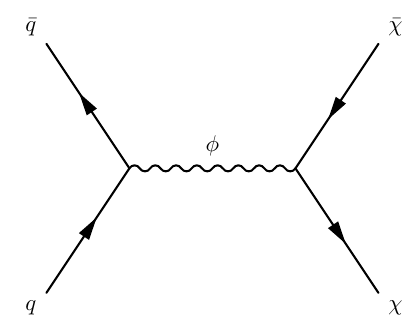}
  \caption{A Feynman diagram showing SIMP pair production via the s-channel exchange of a new scalar mediator.}
  \label{fig:diagram}
\end{figure}

For this search, we assume that the SIMPs are produced in pairs via an $s$-channel exchange of a new scalar mediator that is also coupled to quarks. The Feynman diagram for this process is shown in Fig.~\ref{fig:diagram}. The SIMPs are stable neutral particles that interact with a large cross section with matter but do not hadronize, except by the suppressed higher-order production of quarks via a mediator radiated by one of the SIMP particles. The SIMPs traverse the detector leaving energy in the calorimeters but little activity in the tracking system.  The exact signatures of the resulting trackless jets depend upon the unknown but large interaction cross sections with hadrons and are difficult to predict~\cite{Daci:2015hca}. To perform this search, we adjust the couplings such that the SIMP would be detected as a trackless jet contained completely within the calorimeters. Stronger couplings would give rise to showers starting earlier, \eg in the tracker, and weaker couplings would lead to late extended showers leaking into the muon system. The constrained model under consideration thus provides a framework for exploring the possible pair production of SIMP-induced jets in the CMS calorimeters.

In the analysis presented here, we search for SIMPs yielding trackless jets using a set of $\sqrt{s} = 13 \TeV$ proton-proton (\Pp{}\Pp{}) collision data, corresponding to an integrated luminosity of 16.1\fbinv, collected by the CMS experiment at the LHC in the second half of 2016. In particular, we search for the pair production of SIMPs, and experimentally select the resulting trackless jets using the energy fractions of these jets carried by charged particles (ChF) as a highly effective discriminating observable to suppress the huge QCD multijet background.
In the analysis, we benchmark against a specific model for a SIMP that includes a detailed prescription for its pair production at the LHC and its interaction in the CMS detector. Selection criteria are chosen to optimize the sensitivity for detection of this SIMP, and the results are obtained for this specific model.
Tabulated results are provided in HEPData~\cite{hepdata}.

The ATLAS Collaboration has performed a search~\cite{Aaboud:2019opc} for long-lived neutral particles decaying exclusively in the hadron calorimeter with trackless jets as the experimental signature. However, that search is sensitive to a somewhat different phase space, as in the present search we use a different trigger strategy, and search for a new particle that is seen via its new interactions in both the electromagnetic and hadron calorimeters. The use of a dedicated trigger in the ATLAS analysis, on the one hand makes it possible to lower significantly the jet momentum requirements and, consequently, to boost the sensitivity to trackless jets. On the other hand, jet showers starting in the electromagnetic calorimeter are severely penalized by the event selection, and thus reduced sensitivity is expected for SIMP-nucleon interaction cross  sections at the  level of hadronic cross sections or stronger. The present analysis thus investigates a complementary and poorly explored region of parameter space for new physics.

Noncollider experiments have probed similar phase space as well, considering dark matter masses of order a GeV or less~\cite{Davis:2017noy}. In particular, several direct-detection DM experiments were briefly operated at the Earth's surface~\cite{Angloher:2017sxg,Armengaud:2019kfj}. A direct comparison of these results with collider results, however, depends on the model assumptions~\cite{Hooper:2018bfw,Emken:2018run}.

The paper is organized as follows. Section~\ref{sec:cms} provides a brief description of the CMS detector, and Section~\ref{sec:simulation} presents the SIMP signal model, with a prescription for its simulation. The event reconstruction is described in Section~\ref{sec:reconstruction}. The event selection is given in Section~\ref{sec:event_selection}, and the background estimation, in Section~\ref{sec:background}. Section~\ref{sec:results} then presents the results, and Section~\ref{sec:summary} provides a summary.

\section{The CMS detector} \label{sec:cms}

The central feature of the CMS apparatus is a superconducting solenoid of 6\unit{m} internal diameter, providing a magnetic field of 3.8\unit{T}. Within the solenoid volume are a silicon pixel and strip tracker, a lead tungstate crystal electromagnetic calorimeter (ECAL), and a brass and scintillator hadron calorimeter, each composed of a barrel and two endcap sections. Forward calorimeters extend the pseudorapidity ($\eta$) coverage provided by the barrel and endcap detectors. Muons are detected in gas-ionization chambers embedded in the steel flux-return yoke outside the solenoid. 

Events of interest are selected using a two-tiered trigger system. The first level, composed of custom hardware processors, uses information from the calorimeters and muon detectors to select events at a rate of around 100\unit{kHz} within a fixed latency of about 4\mus~\cite{Sirunyan:2020zal}. The second level, known as the high-level trigger, consists of a farm of processors running a version of the full event reconstruction software optimized for fast processing, and reduces the event rate to around 1\unit{kHz} before data storage~\cite{Khachatryan:2016bia}. 

A more detailed description of the CMS detector, together with a definition of the coordinate system used and the relevant kinematic variables, can be found in Ref.~\cite{Chatrchyan:2008aa}.

\section{Signal simulation} \label{sec:simulation}

The results of the search are compared with predictions based on a specific model for a SIMP, which is described in the following.  Although the comparisons are made against the results of a simulation in the detector of SIMPs specifically as described, the plausibility of some of the model's assumptions is also discussed.

The interaction Lagrangian~(\ref{eq:model}) is implemented in \textsc{FeynRules 2.0}~\cite{Alloul:2013bka} with couplings $g_\chi = -1$ and $g_\cPq = 1$, and mass $m_\phi = 0.14 \GeV$, and interfaced with \MGvATNLO{ v2.1.1}~\cite{Alwall:2014hca} to generate SIMP pair events at leading order using parton distribution function (PDF) set CTEQ6L1. In what follows, the actual choice of the mediator mass is not relevant, as long as off-shell SIMP production is considered.

The SIMP signal is simulated for a range of masses. The lowest mass considered is 1\GeV with a production cross section of $\sigma_{\overline{\chi}\chi} = 15.03 \, \mu \mathrm{b}$, while the highest mass of 1000\GeV has a production cross section of $\sigma_{\overline{\chi}\chi} = 3.63 \unit{fb}$. Using \PYTHIA{ v8.212}~\cite{Sjostrand:2014zea} and tune CUETP8M1~\cite{Khachatryan:2015pea}, we then add an underlying event arising from the fragments of the protons that did not participate in the hard collision, and the generated partons are hadronized. The interactions of the resulting particles with the CMS detector are simulated using \GEANTfour~\cite{Agostinelli:2002hh}, and overlapping \Pp{}\Pp{} collisions (pileup) are overlayed on the main collision.

The interaction of SIMPs with matter is not implemented in \GEANTfour. An implementation of the SIMP interaction Lagrangian as a physics model in \GEANTfour could address this, but is complicated because of possible hadronic physics effects that are not evaluated in the proposed simplified model.

Therefore, since the shower induced by the SIMP interaction is reasonably described by the interaction of a high-momentum neutral hadron, we model the interactions of the SIMPs using neutron-like interactions. However, this description is only approximate, since the neutron is a composite particle that breaks up in the interaction and ceases to exist at high momentum, and may be absorbed at low momentum. By contrast, a SIMP will continue to propagate and induce further interactions and may leave the detector before depositing all its energy.

The assumption of a neutron interaction is only valid for a certain range of couplings. As described in Ref.~\cite{Daci:2015hca}, decreasing the SIMP-nucleon interaction cross section $\sigma_{\chi, \mathrm{N}}\sim g_\cPq^2 g_{\chi}^2$ by a factor 10 reduces the signal acceptance by a factor 6. An increase in cross section, on the other hand, is constrained from above by measurements of the cosmic microwave background~\cite{Daci:2015hca}. Our assumption of a hadron-like interaction cross section with the detector material is thus a reasonable choice to demonstrate the experimental signature targeted with the simplified model considered.

\begin{figure}[h!]
  \centering
  \includegraphics[width=0.49\textwidth]{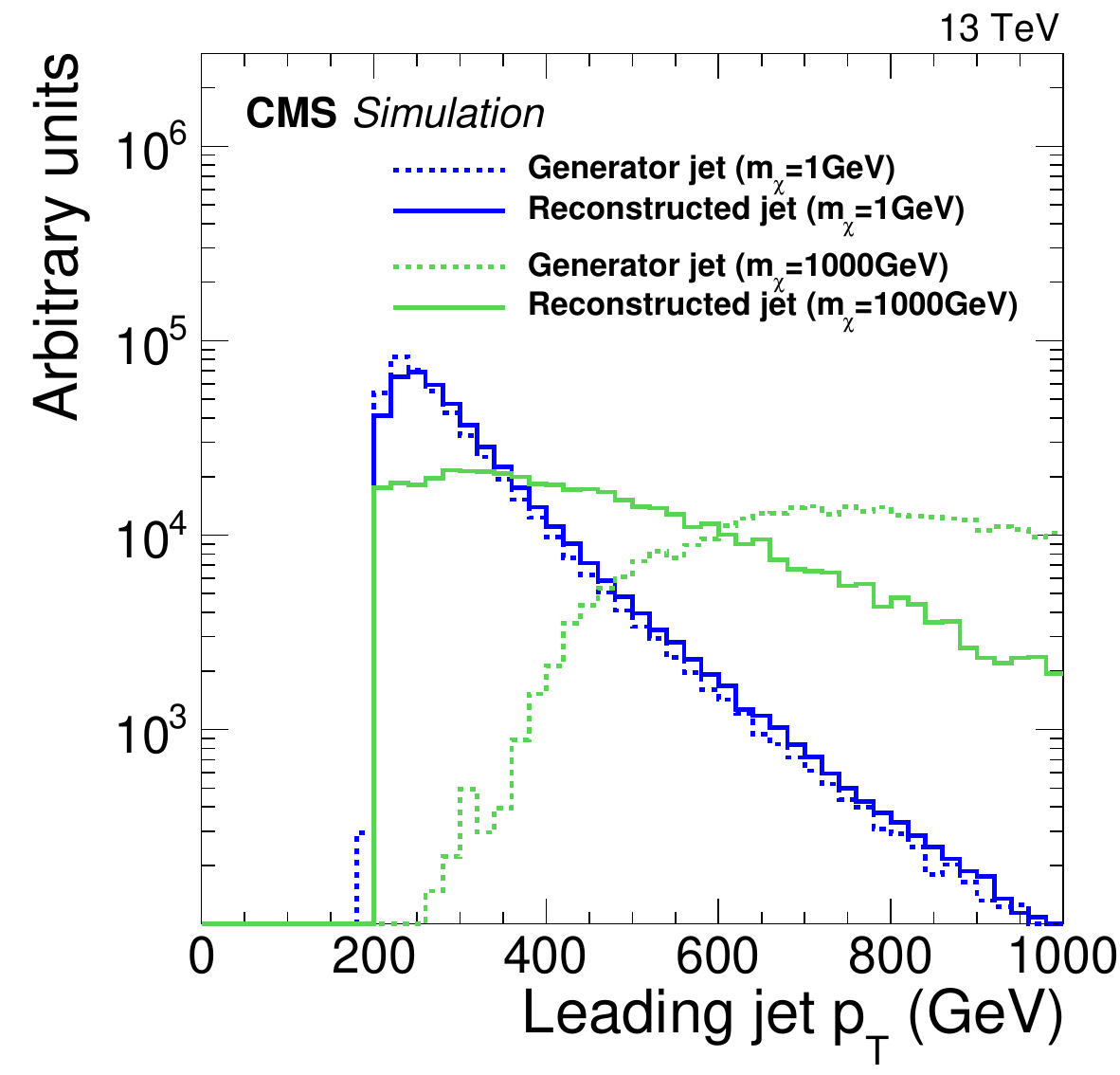}
  \includegraphics[width=0.49\textwidth]{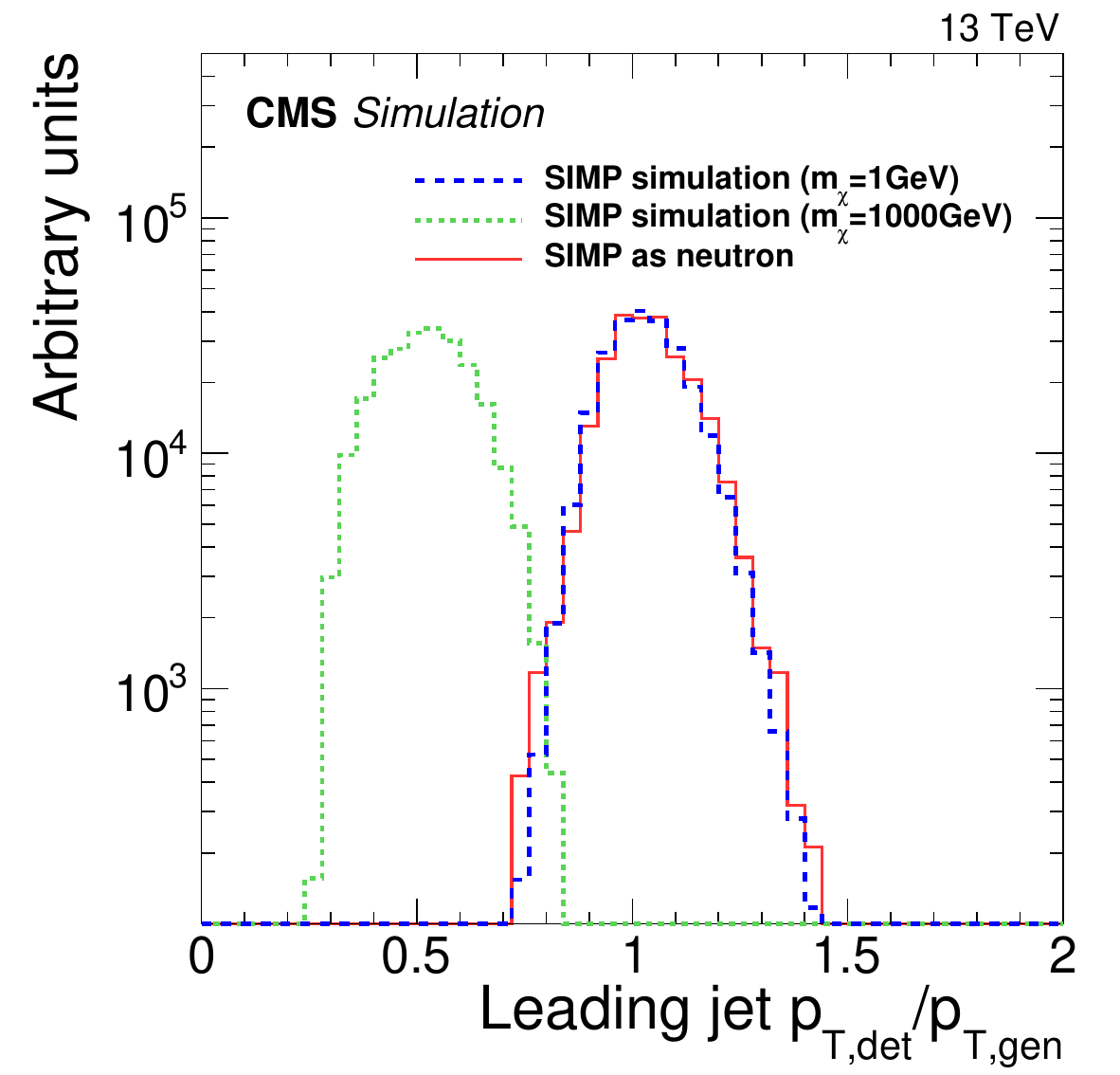}
  \caption{Left: comparison of the leading jet $\pt$ spectrum between jets clustered at the generator level (dashed lines) and after detector simulation (solid lines), arising from the \GEANTfour SIMP simulation at masses of 1\GeV (blue lines) and 1000\GeV (green lines); \cmsRight: the ratio of the reconstruction-level and generator-level jet transverse momenta for the \GEANTfour SIMP simulation at masses of 1\GeV (dark blue, long-dashed line) and 1000\GeV (green, short-dashed line), including a comparison to a simulation where a $1\GeV$ SIMP is replaced by a neutron (red, solid line).}
  \label{fig:g4simp}
\end{figure}

To implement this neutron-like interaction, we added to the \GEANTfour simulation a new SIMP particle as a clone of the neutron, but with an adjustable mass. The SIMP was set to deposit only its kinetic energy in the interaction, and not its mass.
To simplify the setup, we use only the inelastic part of the neutron interaction, which dominates at high momentum. As a further approximation, we consider only the first SIMP interaction in simulation. Since a true SIMP could undergo additional interactions before leaving the calorimeter, our approach of including only the first interaction conservatively represents an underestimate of the observable energy in the induced shower.
With this setup, SIMP signal samples are simulated and reconstructed (as discussed in Section~\ref{sec:reconstruction}), and narrow jets with large neutral hadron energy fractions are obtained.
In Fig.~\ref{fig:g4simp} (\cmsLeft), we compare the transverse momentum of the leading jet with $\pt>200\GeV$ at the generator level (\ie~carrying the full SIMP momentum) with the momentum of the corresponding reconstructed jet. In Fig.~\ref{fig:g4simp} (\cmsRight), we show the ratio of these reconstructed and generated transverse momenta, including the comparison to the case where a SIMP of mass 1\GeV is replaced by a neutron. The latter comparison verifies that this SIMP simulation matches that obtained with neutrons in the standard version of \GEANTfour in the phase space relevant for the analysis.

Figure~\ref{fig:g4simp} illustrates that while SIMPs with large incident momenta and with the mass of a neutron will deposit virtually all their momentum in the first interaction~\cite{Daci:2015hca}, high-mass SIMPs will transfer only a part of their momentum in collisions with the low-mass nucleons at rest in the detector material, and will thus induce smaller shower energy depositions than if they had a small mass.
Using our simulation setup, we indeed observe a suppression of the reconstructed jet energies due to reduced shower depositions.
As an example, we see that a SIMP with a 1000\GeV mass and $\pt > 200 \GeV$ leads on average to a jet momentum about half as large as for a neutron of the same momentum. However, from kinematic considerations in elastic scattering, a significantly smaller momentum transfer may be expected at such high SIMP mass, depending on the target mass. The approach in the simulation of this trackless-jet test model, of treating the SIMP as having neutron-like interactions at all masses, must thus be seen as an approximative assumption. Allowing coherent scattering of the SIMP off the calorimeter nuclei, the interactions of the SIMPs with mass up to about 100\GeV, \ie of the order of the mass of those nuclei, are expected to be kinematically well modelled. At higher masses, the simulation is more exploratory and is presented as a yardstick in the current absence of a more developed treatment of the SIMP-nucleon interaction.

\section{Event reconstruction} \label{sec:reconstruction}

In this analysis, we search for jet-like objects with very small ChF values. To reconstruct and identify these objects, we take as input the charged and neutral hadrons, photons, electrons, and muons, all of which are coherently reconstructed by a particle-flow event algorithm~\cite{Sirunyan:2017ulk}. Charged hadrons not associated to the primary interaction vertex are removed to mitigate the effect of pileup collisions. Next, we cluster these particles into jets using the anti-\kt algorithm~\cite{Cacciari:2008gp, Cacciari:2011ma} with a distance parameter $R=0.4$, which by construction provides an unambiguous association of tracks with jets.
The charged fraction ChF of a jet is then defined as the ratio of the scalar sum of the transverse momenta of all charged hadrons associated with the jet to the transverse momentum of the jet itself.
The energies of these jets are subsequently further corrected for contributions from pileup and for $\eta$- and \pt-dependent response biases~\cite{Khachatryan:2016kdb}.

{\tolerance=800 The candidate vertex with the largest value of summed physics-object $\pt^2$ is taken to be the primary \Pp{}\Pp{} interaction vertex. The physics objects are the jets, clustered using the jet finding algorithm~\cite{Cacciari:2008gp,Cacciari:2011ma} with the tracks assigned to candidate vertices as inputs, and the associated missing \pt, taken as the negative vector sum of the \pt of those jets. \par}

Since the principal discriminant for identifying SIMP candidates is ChF, it is important to minimize incorrect primary vertex identification, because the removal of spurious charged hadron tracks originating from pileup depends on their primary vertex association. While jets with many high-momentum tracks can usually be associated with a primary vertex, this is not the case for the neutral jets of signal events. For these jets, the underlying event and initial-state QCD radiation may provide some tracks, but it is likely that the wrong vertex is selected. In such cases, the removal of charged particles not associated with the chosen primary vertex also removes the tracks from the SIMP production vertex. However, an incorrect choice of vertex in signal events has little effect as their jets exhibit a low ChF already.

The correct choice of the primary vertex is made in well above 99\% of QCD multijet background events. However, if the primary vertex is wrongly chosen in these background events, the pileup suppression procedure may purge tracks from the true vertex, resulting in the spurious appearance of neutral jets. This makes such an event appear signal-like.
For the most stringent ChF requirements considered in this analysis, this reconstruction-induced background becomes dominant as compared with backgrounds from prompt photons and very rare jet fragmentation into mostly neutral hadrons and photons.

Simulation studies on background events have shown that in the very rare case when the first vertex is wrongly chosen, the second of the $\pt^2$-ordered list of reconstructed vertices is the true vertex from the hard collision in more than 50\% of the cases, often because a single poor-quality track from a pileup collision is erroneously reconstructed with high momentum.
Therefore, to mitigate this reconstruction-induced background, we reconstruct each event twice: once with the standard reconstruction, and again, assuming the second vertex to be the collision vertex. In the case that the second vertex is the correct one, QCD jets acquire larger values of ChF compared to those obtained with the default reconstruction. Thus the subsequent event selection requires the condition set on ChF to be satisfied for both vertex choices.
Additional background suppression using lower-ranked primary vertices was found not to further improve the sensitivity of the signal selection.

Since photons are reconstructed as neutral jets, we need to efficiently identify and reject them. In this analysis, we identify photons using loose identification requirements~\cite{Khachatryan:2015iwa}. To further increase the photon identification efficiency, we also consider as photons those jets not coinciding with a loose photon but containing a reconstructed electron-positron pair (potentially coming from photon conversion) whose \pt is greater than 30\% of that of the jet itself.

\section{Event selection} \label{sec:event_selection}

This analysis used only a portion of the data collected during 2016 because, for the early part of that running period, saturation-induced dead time was present in the readout of the silicon strip tracker. This caused hard-to-model instantaneous-luminosity-dependent inefficiencies for the reconstruction of tracks, which led to subtle event-wide correlations that prevented a reliable prediction of the background arising from low-charge jets in QCD multijet events. With this detector issue corrected for the second half of 2016, a dataset was collected corresponding to an integrated luminosity of 16.1\fbinv, and events passing an online selection (trigger) algorithm requiring a jet with $\pt > 450 \GeV$ were used for this analysis.

As a baseline offline selection, we select two jets, each with $\pt > 550 \GeV$, such that the applied trigger requirements are 98\% efficient for the selected events. Furthermore, we require these jets to have $\abs{\eta} < 2.0$, so they are fully within the tracking volume, thus suppressing backgrounds from jets that have tracks falling outside of the tracker acceptance, resulting in an underestimation of ChF.

Except for the suppressed process of SIMPs radiating a mediator that decays into quarks, SIMPs do not undergo parton showering themselves, while quarks and gluons undergo QCD final-state radiation. Therefore, events with SIMPs have on average a lower number of jets compared with QCD multijet background events. To suppress this background, we reject events if in addition to the two already selected jets other jets are found with $\pt>30 \GeV$ and $\abs{\eta}<5$. The same radiation argument also implies that the selected high-\pt jets are better separated in azimuth in signal events than in QCD multijet background events. Following this, we further require an azimuthal separation of $\Delta\phi > 2$ between the two selected jets.

We also apply a photon veto to suppress $\PGg$+jets events. This is done by rejecting events for which the identified photon with the highest \pt falls within $\Delta R = \sqrt{\smash[b]{(\Delta\eta)^2+(\Delta\phi)^2}} < 0.1$ of the leading or subleading jet. In cases where the electromagnetic energy fraction of the jet carried by neutral particles is larger than 0.8, but the photon candidate in the jet does not satisfy the identification requirements, we still reject the event in the case that a conversion is found within $\Delta R < 0.2$ of the photon candidate, as described in Section~\ref{sec:reconstruction}. Furthermore, the photon veto is complemented by requiring both jets to have an electromagnetic energy fraction carried by neutral particles lower than 0.9, which additionally removes spurious jets formed around anomalous ECAL deposits.
Finally, we apply a dedicated selection~\cite{Sirunyan:2019kia} to remove beam halo events.

Because standard jet identification criteria would suppress the trackless jets of our signal process and cannot be applied in this analysis, it may be possible for a spurious jet to pass our selection criteria. However, because the rate of two simultaneous, independent signals of high-energy calorimeter noise is insignificant, the probability to select two back-to-back high-\pt noise jets is negligible. In addition, we verify that individual events with jets at the smallest ChF values do not exhibit unexpected features, using both the QCD multijet simulation, and events in the triggered data sample that do not pass the jet \pt thresholds.

In the following, we refer to the sample of events satisfying the above set of selection criteria as the ``baseline selection''.

\begin{figure}[h]
  \centering
  \includegraphics[width=0.49\textwidth]{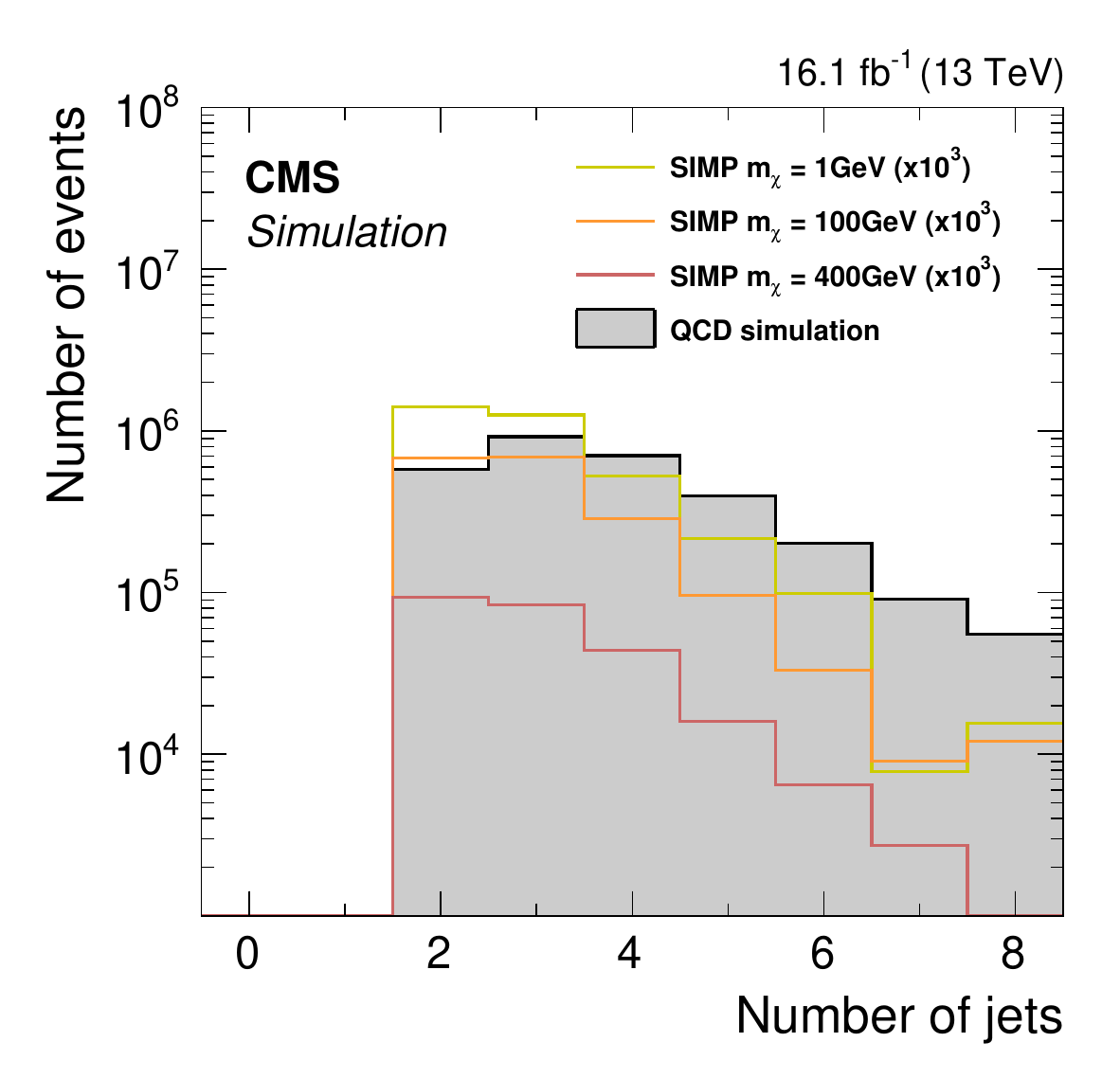}
  \includegraphics[width=0.49\textwidth]{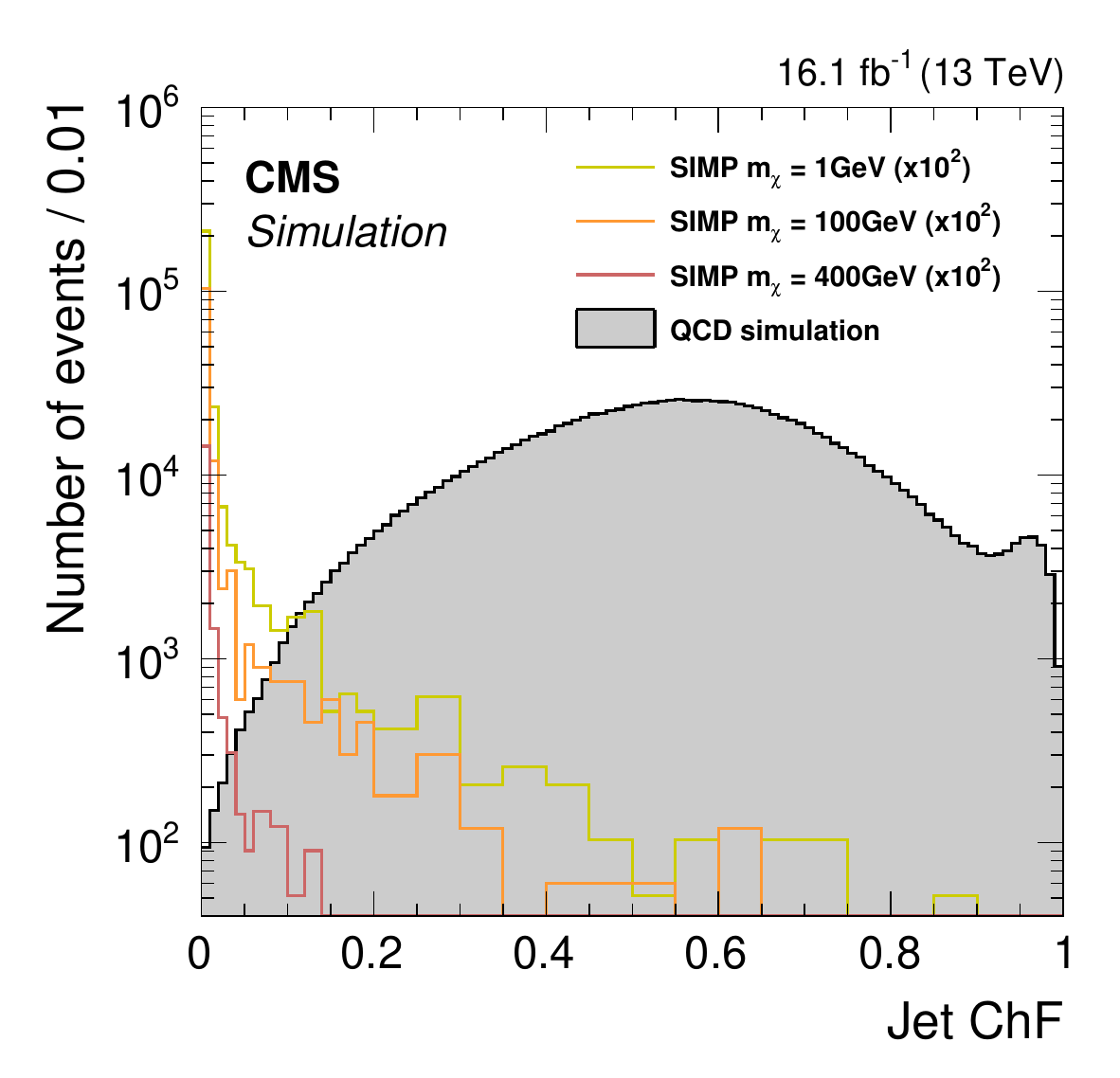}
  \caption{Distributions of the number of jets with $\pt>30 \GeV$ and $\abs{\eta}<5$ (\cmsLeft), and the value of ChF of the two leading jets (\cmsRight). The simulated QCD multijet background is compared with the signal expected for three different SIMP masses, with their cross sections scaled as indicated in the legend. The baseline selection is applied, except the events with three or more jets with $\pt>30 \GeV$ and $\abs{\eta}<5$ are included in the number of jets in the left plot.}
  \label{fig:event_selection}
\end{figure}

Figure~\ref{fig:event_selection} (\cmsLeft) shows the distribution of the number of jets for simulated events satisfying the baseline selection criteria, except for the rejection of events with three or more jets with $\pt>30 \GeV$ and $\abs{\eta}<5$. Figure~\ref{fig:event_selection} (\cmsRight) depicts the distribution of ChF values for the two leading jets for simulated events satisfying the baseline selection criteria. The predicted QCD multijet background is compared with the signal expected for three different SIMP masses. The ChF distributions for QCD simulation and signal are very different, with the signal peaking strongly at low ChF, just where the QCD events are minimal.

In order to estimate the QCD multijet background from data, we define a control region consisting of a subsample of events satisfying the baseline selection, where at least one of the two leading jets has ChF greater than 0.25. For this control sample selection, we apply the ChF requirement only to jets reconstructed using the default primary vertex. The presence of at least one jet with a large value of ChF ensures that the correct primary vertex is selected.

Candidate signal events are selected from the baseline event selection by requiring both leading jets to have ChF below a certain threshold, both for the default and for the alternate choice of the primary vertex.

\section{Background estimation} \label{sec:background}

The $\PGg$+jets background is shown to be insignificant, as no events remain after the event selection is applied to a simulated sample corresponding to an integrated luminosity of 27\fbinv. The associated uncertainty is smaller than any of the other systematic uncertainties in the estimation of the total background.

The main QCD multijet background is simulated using \MGvATNLO{ v2.2.2} at leading order using PDF set NNPDF 3.0,
with the \PYTHIA{ v8.212} tune CUETP8M1 for the underlying event. Interactions in the detector are simulated with \GEANTfour, and pileup collisions are overlayed.

The QCD multijet background is not described accurately by the simulation, especially at low ChF. The differences between data and simulation are not problematic, since we estimate the QCD multijet background from data, while using simulated events only to validate the background estimation procedure.

As a first step, we measure the ChF selection efficiency of jets in the control sample by picking one jet with large ChF ($>$0.25) and applying the ChF selection on the other jet. This measurement is done in 6 bins of jet \pt and 8 bins of jet $\eta$. The number of QCD events in the signal region is then estimated using the QCD dijet events passing the baseline selection requirements described in Section~\ref{sec:event_selection}. For each such event, we use the previously measured ChF selection efficiencies corresponding to the \pt and $\eta$ of the two leading jets as two independent weights multiplied to obtain a weight by which the considered event enters the background prediction (2-leg prediction). Alternatively, events with one jet with ChF below the signal requirement can be used, where the measured efficiencies are then applied on the other jet (1-leg prediction).

As a first check, a closure test is performed on the background prediction method using jets clustered from particles at the generator level, before interaction with the detector. Agreement within statistical uncertainties between the generator-level expectation and the 1- and 2-leg predictions confirms that no relevant underlying physical correlations are present between the two jets, and also confirms that the choice of \pt and $\eta$ bin sizes of the ChF efficiencies is adequate.

\begin{figure}[h]
  \centering
  \includegraphics[width=0.49\textwidth]{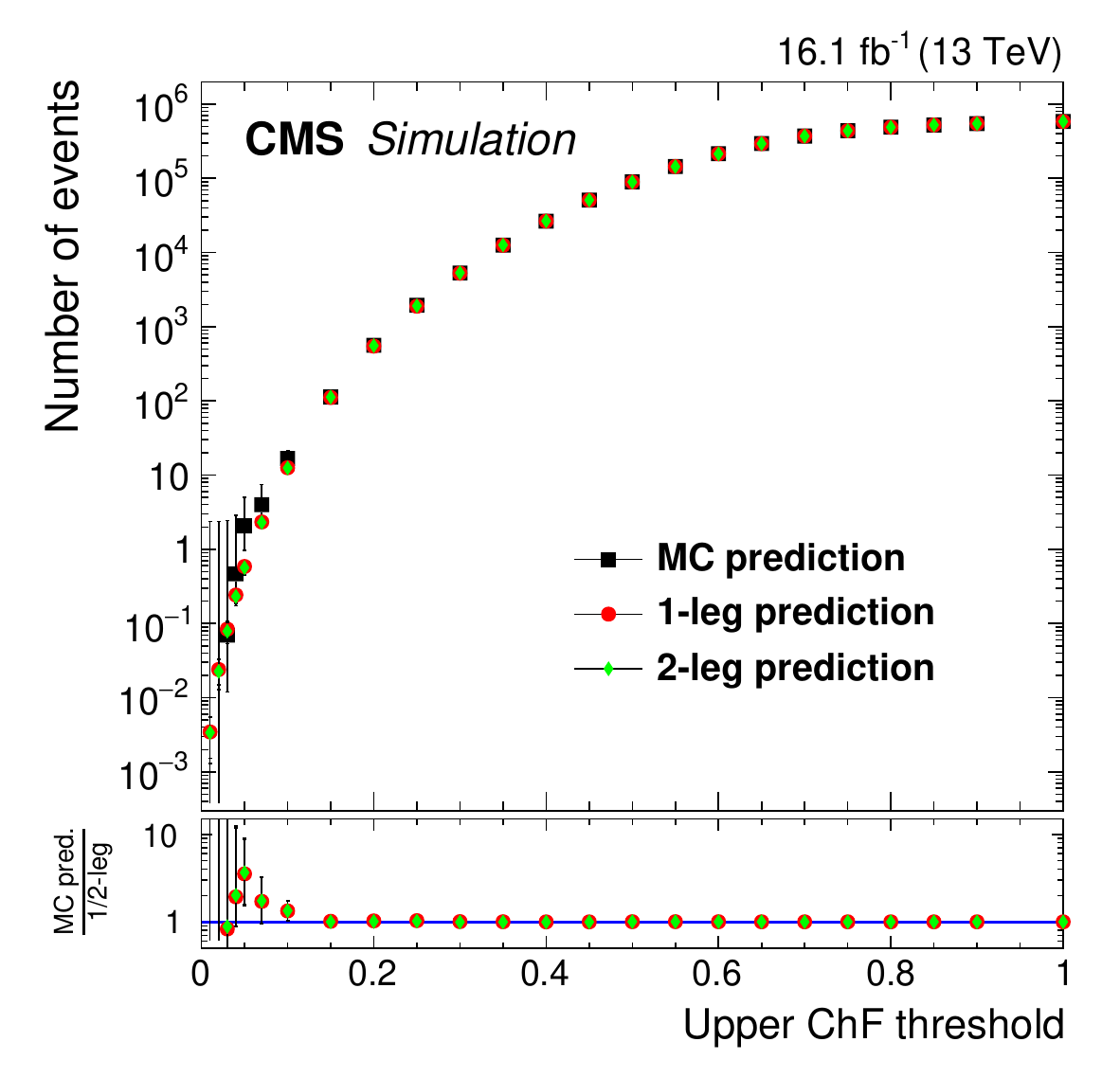}
  \caption{The number of background events obtained from the 1- and 2-leg predictions using reconstructed objects in simulation, compared to the direct prediction from MC simulation, shown for various upper ChF thresholds. The bottom panel shows the ratios of the MC prediction to the 1-leg and the 2-leg background predictions.}
  \label{fig:closuretest}
\end{figure}

{\tolerance=800 A further closure test is done by using the simulation as the data sample, and comparing the Monte Carlo (MC) expectation with the 1- and 2-leg predictions using reconstructed objects in simulation, as shown in Fig.~\ref{fig:closuretest}. For the MC expectation, the ChF selection is applied to the two leading reconstructed jets, for both choices of the primary vertex. As can be seen from the plot, the method correctly predicts the multijet background within the statistical precision of the test, proving that no significant correlations between the jets are introduced by the event reconstruction. The systematic uncertainty in the background estimate is taken to be the statistical uncertainty of the test or the difference between the generator-level information and the prediction, whichever is the larger. This uncertainty becomes dominant for lower ChF thresholds and reaches up to 250\% for $\mathrm{ChF} < 0.05$. \par}

\begin{figure}[h]
  \centering
  \includegraphics[width=0.49\textwidth]{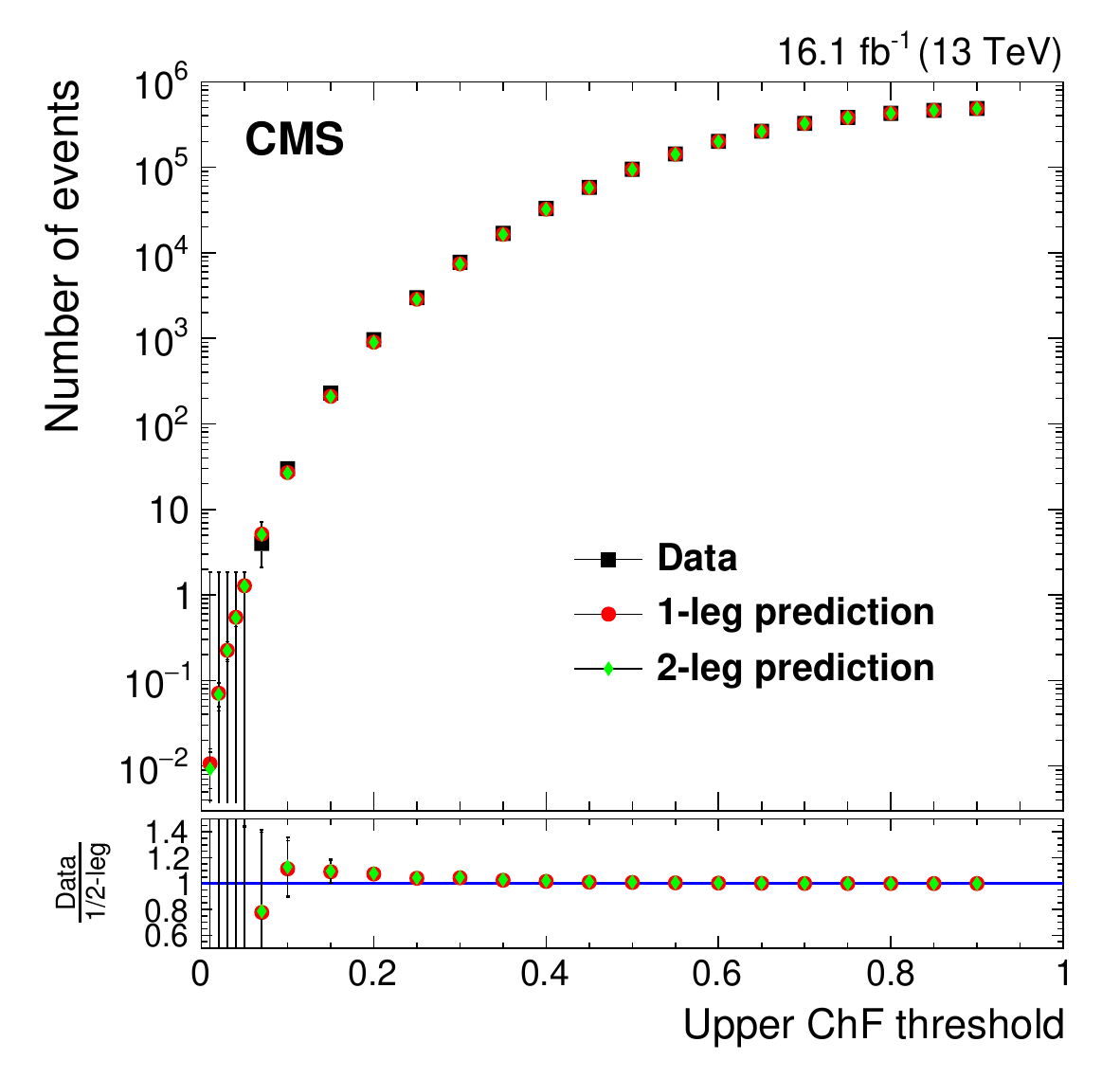}
  \caption{The number of background events obtained from the 1- and 2-leg predictions derived from data, together with the direct observation in data, in bins in ChF, where either the leading or subleading jet has a ChF within the bin edges, and both have a ChF below the upper bin threshold. The bottom panel shows the ratios of the observation in data to the 1-leg and the 2-leg background predictions.}
  \label{fig:prediction}
\end{figure}

Next, we predict the background using data and compare with the observed data. To demonstrate the closure of the method without potential contamination from a signal at low ChF, this comparison is done using bins where either the leading or the subleading jet has a ChF within the bin edges, and both jets have a ChF below the upper threshold of the bin. This comparison is shown in Fig.~\ref{fig:prediction}. The 1- and 2-leg predictions agree within uncertainties in data, confirming that no correlations between the jets are present. The agreement demonstrates a reliable prediction of the bulk of the ChF distribution and the normalization of the background.

{\tolerance=800 Apart from the physical sources of photon and QCD multijet background, other sources of an instrumental or algorithmic nature may arise, \eg the previously mentioned possibility of incorrectly choosing the primary vertex. To ensure the background prediction method does not underestimate such additional sources of background, detailed checks were performed using the events with the lowest ChF jets from the QCD multijet simulation, as well as in a slightly larger data sample of events collected using the same online trigger, but which did not pass the offline jet \pt requirements. During these checks, no anomalous events were observed satisfying the baseline event selection. \par}

\section{Results} \label{sec:results}

Table~\ref{tab:results} shows the number of predicted and observed events, along with the expected yield from a SIMP signal for three different SIMP masses, for various values of the ChF requirement. 
The background prediction is obtained using the 2-leg prediction, since it has a nearly identical statistical uncertainty to the 1-leg prediction but avoids the nontrivial statistical overlap between the event sample used to measure the binned efficiencies, and the sample to which these efficiencies are applied to obtain the background prediction. The systematic uncertainty in the data prediction is dominated by the previously described uncertainty related to the closure test. Additionally, a statistical uncertainty of up to 17\% arising from the measured efficiencies of the ChF selection is accounted for, as is a 2\% inefficiency of the trigger observed after the offline jet \pt requirement of 550\GeV. 

\renewcommand{\arraystretch}{1.3}
\begin{table*}[ht]
  \centering
\topcaption{The numbers of background and observed events for different upper bounds on the ChF value. The background estimations are derived using the data-based 2-leg predictions. The expected number of signal events is given for the $m_{\chi} = 1$, 100, and 1000\GeV scenarios, with the corresponding statistical uncertainties.}
\begin{tabular}{c r@{$\,$}r@{$\,$}r@{$\,$}r@{$\,$}r r r@{$\,$}r@{$\,$}r r@{$\,$}r@{$\,$}l r@{$\,$}r@{$\,$}r}
\hline
\raisebox{-1mm}{ChF selection} & \multicolumn{5}{c}{\raisebox{-1mm}{Background prediction}} & \multirow{2}{*}{Obs.} & \multicolumn{9}{c}{SIMP signal [$m_{\chi}$]} \\
\raisebox{1mm}{criterion} & \multicolumn{5}{c}{\raisebox{1mm}{from data}} & & \multicolumn{3}{c}{1\GeV} & \multicolumn{3}{c}{100\GeV} & \multicolumn{3}{c}{1000\GeV}\\
\hline
$<$0.20 & 898 & $\pm$ & 30\stat & $\pm$ & 33\syst & 969 & 1300 & $\pm$ & 58 & 634 & $\pm$ & 44 & 2.25 & $\pm$ & 0.07 \\
$<$0.15 & 209 & $\pm$ & 10\stat & $\pm$ & 17\syst & 229 & 1269 & $\pm$ & 57 & 613 & $\pm$ & 43  & 2.18 & $\pm$ & 0.07  \\
$<$0.10 & 26.6 & $\pm$ & 2.2\stat & $\pm$ & 9.3\syst & 30 & 1197 & $\pm$ & 56 & 589 & $\pm$ & 42  & 2.09 & $\pm$ & 0.07 \\
$<$0.07 & 5.1 & $\pm$ & 0.6\stat & $\pm$ & 4.1\syst & 4 & 1153 & $\pm$ & 55 & 568 & $\pm$ & 41  & 2.00 & $\pm$ & 0.07 \\
$<$0.05 & 1.27 & $\pm$ & 0.22\stat & \multicolumn{2}{l}{$\!\!\!\!^{+\ 3.40}_{-\ 1.27\,}$\syst} & 0 & 1101 & $\pm$ & 53 & 544 & $\pm$ & 40  & 1.90 & $\pm$ & 0.06 \\
\hline
\end{tabular}
\label{tab:results}
\end{table*}

The signal region used to determine the final results is defined by $\mathrm{ChF} < 0.05$. This rejects most of the QCD background, while avoiding tighter ChF requirements, where the generator-level information used in the closure tests starts to yield large statistical uncertainties, and where higher-order contributions from mediator radiation off the SIMPs could become nonnegligible.

Using these results, we calculate model-independent limits at 95\% confidence level (\CL) using the \CLs criterion with a profile likelihood modified for upper limits as test statistics, in which the systematic uncertainties are modelled as nuisance parameters~\cite{CLS2,CLS1}. All included systematic uncertainties are profiled with a lognormal constraint, except for the uncertainty in the background estimation, which is dominated by the statistical uncertainty associated with the closure test, and is profiled with a gamma function. This results in both an observed and an expected visible cross section upper limit of $\sigma_{\mathrm{vis}}^{95\%} = \sigma\,\mathrm{A}\,\epsilon =  0.18\ \unit{fb}$, with $\mathrm{A}$ the acceptance and $\epsilon$ the event selection efficiency.

For the SIMP signals, as is done for data events, the event selection requirements are applied to jets for both primary vertex choices. The 95\% \CL upper limits on the SIMP production cross section are then calculated for SIMP masses between 1 and 1000\GeV, for the signal region with $\mathrm{ChF} < 0.05$, using the same procedure as described for the model-independent limit. 

Several systematic uncertainties are assigned to the estimation of the signal. Uncertainties arising from the jet energy corrections are evaluated assuming the jets to be clustered from calorimetric input only, and range from 2.8 to 6.3\%, increasing with decreasing SIMP mass. Furthermore, uncertainties related to the integrated luminosity~(2.5\%)~\cite{CMS-PAS-LUM-17-001}, to the trigger efficiency mentioned before in the context of the background (2\%), and to the limited signal sample size (2.9 to 7.4\%) are included. Other potential experimental sources of uncertainty, like the photon and conversion veto requirements and the  effect of pileup, are found to be negligible.

The results are compared with the predictions of a specific model for the production of SIMPs at the LHC and for the SIMPs interactions in the CMS detector.  This model is described in Section~\ref{sec:simulation}, where its relevance for potential SIMPs is also discussed.  The results are benchmarked against a specific model implementation and therefore no modelling uncertainties are incorporated into the analysis. Uncertainties related to the simulation of the simplified theoretical model, \eg uncertainties arising from scale assumptions or PDFs, are not included, as selection acceptance uncertainties arising from these sources were found to be negligible.

\begin{figure}[h]
  \centering
  \includegraphics[width=0.5\textwidth]{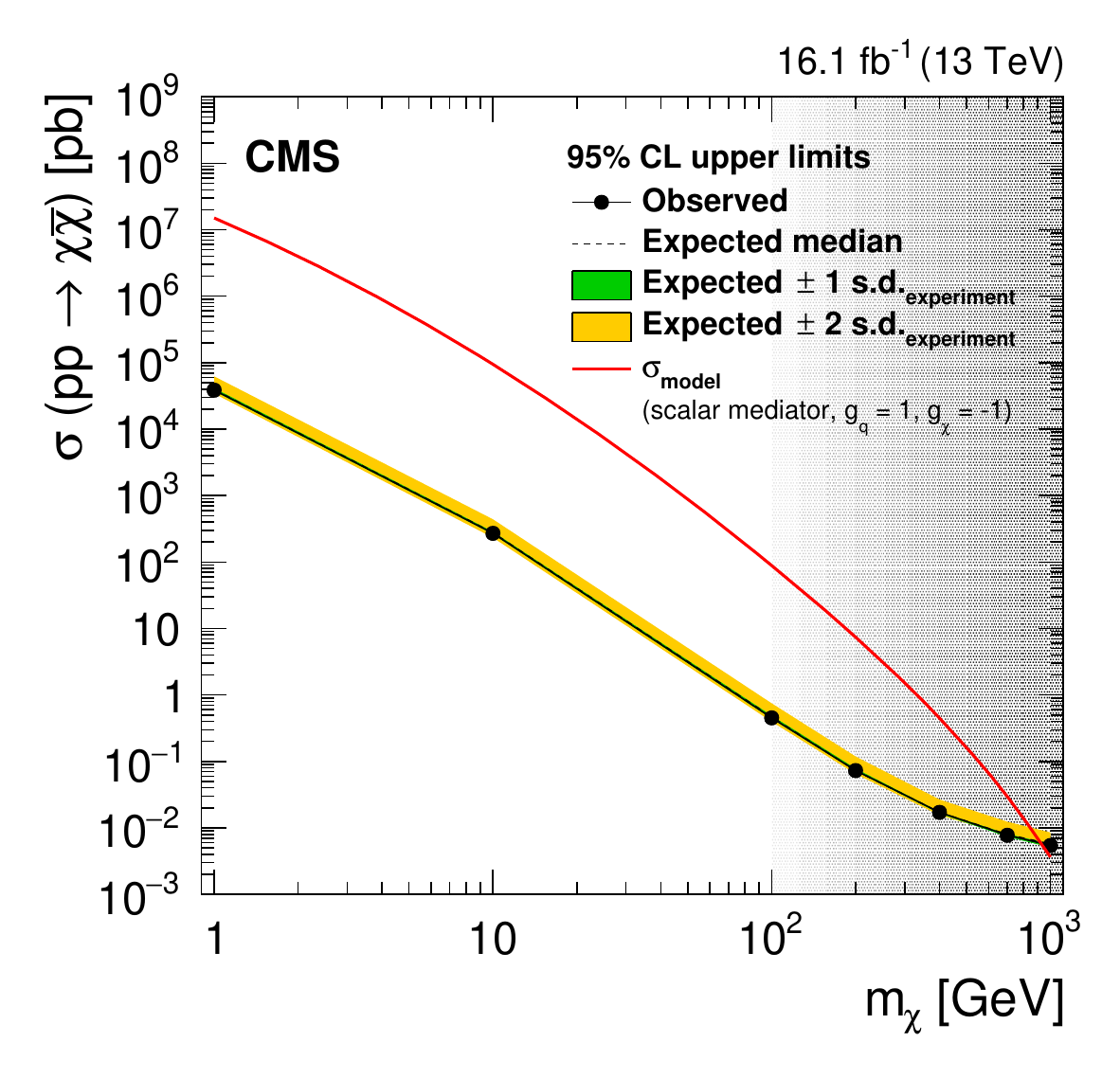}
  \caption{The expected and observed 95\% \CL upper limits on the production cross section for SIMPs with masses between 1 and 1000\GeV, with the assumption that the SIMP interaction in the detector can be approximated as neutron-like. The theoretical prediction of a simplified model incorporating this approximation and including a scalar mediator with couplings $g_\chi = -1$ and $g_\cPq = 1$ is also shown (red line). For masses above 100\GeV, where the modelling of the SIMP-nucleon interaction becomes more speculative, the obtained cross section upper limits are increasingly uncertain (shaded area).}
  \label{fig:SIMP_limit_ChF0p1}
\end{figure}

{\tolerance=800 Figure~\ref{fig:SIMP_limit_ChF0p1} shows the  expected and observed 95\% CL upper limits on the production cross section for SIMPs with masses between 1 and 1000\GeV.  These limits are obtained for off-shell production of the SIMP pair through a new scalar mediator with couplings $g_\chi = -1$ and $g_\cPq = 1$, under the assumption that the SIMP's interaction in the detector is neutron-like, as described in Section~\ref{sec:simulation}. Within this framework, we exclude SIMP masses up to 100\GeV, which includes the phenomenologically most interesting low-mass phase space~\cite{Daci:2015hca}. At higher masses, the limits shown are subject to the caveats discussed in Section~\ref{sec:simulation}. \par}

In the case of production through a new vector mediator, the production cross section is between 15 and 30\% larger~\cite{Daci:2015hca} compared to the scalar mediator that is assumed here.

The results of this search are interpreted in terms of a specific benchmark model. What is needed to go beyond this benchmark is a fully developed theoretical prediction of a SIMP that provides a framework for understanding the interactions of a SIMP with matter. Given this understanding, the results of the present search could be further interpreted as limits for such a SIMP, and the relevance of the benchmark assumptions for the theory could be determined.

\section{Summary} \label{sec:summary}

{\tolerance=800 A search has been presented for dark matter in the form of strongly interacting massive particles (SIMPs) manifesting themselves in the detector as trackless jets. The large multijet background is efficiently suppressed using the charged energy fraction of jets as the key discriminator. The remaining background is estimated directly from data. Using proton--proton collision data corresponding to an integrated luminosity of 16.1\fbinv  collected by the CMS experiment in 2016, we set first limits on the production cross section for SIMPs with masses between 1 and 1000\GeV at 95\% confidence level (\CL), using a signal simulation that assumes the SIMP interaction in the detector can be approximated as neutron-like. The signal modelling assumptions stated previously have small uncertainties, and hence a small impact on the cross section upper limits, for SIMP masses up to about 100\GeV, but become increasingly uncertain above 100\GeV, where an improved phenomenology of the SIMP-nucleon interaction would be welcome. Within this framework we exclude SIMPs with masses less than 100\GeV. These limits were obtained for the off-shell production of SIMP pairs, through a new scalar mediator with couplings $g_\chi = -1$ and $g_\cPq = 1$. An upper limit on the fiducial cross section of $0.18\unit{fb}$ at 95\% \CL is also provided for a generic signal of high-momentum trackless jets. With this search, strongly interacting massive particles, for which the interaction strength is constrained to generate a trackless jets signature, have been ruled out over a wide mass range. \par}

\begin{acknowledgments}

  We congratulate our colleagues in the CERN accelerator departments for the excellent performance of the LHC and thank the technical and administrative staffs at CERN and at other CMS institutes for their contributions to the success of the CMS effort. In addition, we gratefully acknowledge the computing centres and personnel of the Worldwide LHC Computing Grid and other centres for delivering so effectively the computing infrastructure essential to our analyses. Finally, we acknowledge the enduring support for the construction and operation of the LHC, the CMS detector, and the supporting computing infrastructure provided by the following funding agencies: BMBWF and FWF (Austria); FNRS and FWO (Belgium); CNPq, CAPES, FAPERJ, FAPERGS, and FAPESP (Brazil); MES (Bulgaria); CERN; CAS, MoST, and NSFC (China); MINCIENCIAS (Colombia); MSES and CSF (Croatia); RIF (Cyprus); SENESCYT (Ecuador); MoER, ERC PUT and ERDF (Estonia); Academy of Finland, MEC, and HIP (Finland); CEA and CNRS/IN2P3 (France); BMBF, DFG, and HGF (Germany); GSRT (Greece); NKFIA (Hungary); DAE and DST (India); IPM (Iran); SFI (Ireland); INFN (Italy); MSIP and NRF (Republic of Korea); MES (Latvia); LAS (Lithuania); MOE and UM (Malaysia); BUAP, CINVESTAV, CONACYT, LNS, SEP, and UASLP-FAI (Mexico); MOS (Montenegro); MBIE (New Zealand); PAEC (Pakistan); MSHE and NSC (Poland); FCT (Portugal); JINR (Dubna); MON, RosAtom, RAS, RFBR, and NRC KI (Russia); MESTD (Serbia); SEIDI, CPAN, PCTI, and FEDER (Spain); MOSTR (Sri Lanka); Swiss Funding Agencies (Switzerland); MST (Taipei); ThEPCenter, IPST, STAR, and NSTDA (Thailand); TUBITAK and TAEK (Turkey); NASU (Ukraine); STFC (United Kingdom); DOE and NSF (USA).
  
  \hyphenation{Rachada-pisek} Individuals have received support from the Marie-Curie programme and the European Research Council and Horizon 2020 Grant, contract Nos.\ 675440, 724704, 752730, 765710 and 824093 (European Union); the Leventis Foundation; the Alfred P.\ Sloan Foundation; the Alexander von Humboldt Foundation; the Belgian Federal Science Policy Office; the Fonds pour la Formation \`a la Recherche dans l'Industrie et dans l'Agriculture (FRIA-Belgium); the Agentschap voor Innovatie door Wetenschap en Technologie (IWT-Belgium); the F.R.S.-FNRS and FWO (Belgium) under the ``Excellence of Science -- EOS" -- be.h project n.\ 30820817; the Beijing Municipal Science \& Technology Commission, No. Z191100007219010; the Ministry of Education, Youth and Sports (MEYS) of the Czech Republic; the Deutsche Forschungsgemeinschaft (DFG), under Germany's Excellence Strategy -- EXC 2121 ``Quantum Universe" -- 390833306, and under project number 400140256 - GRK2497; the Lend\"ulet (``Momentum") Programme and the J\'anos Bolyai Research Scholarship of the Hungarian Academy of Sciences, the New National Excellence Program \'UNKP, the NKFIA research grants 123842, 123959, 124845, 124850, 125105, 128713, 128786, and 129058 (Hungary); the Council of Science and Industrial Research, India; the Latvian Council of Science; the Ministry of Science and Higher Education and the National Science Center, contracts Opus 2014/15/B/ST2/03998 and 2015/19/B/ST2/02861 (Poland); the National Priorities Research Program by Qatar National Research Fund; the Ministry of Science and Higher Education, project no. 0723-2020-0041 (Russia); the Programa Estatal de Fomento de la Investigaci{\'o}n Cient{\'i}fica y T{\'e}cnica de Excelencia Mar\'{\i}a de Maeztu, grant MDM-2015-0509 and the Programa Severo Ochoa del Principado de Asturias; the Thalis and Aristeia programmes cofinanced by EU-ESF and the Greek NSRF; the Rachadapisek Sompot Fund for Postdoctoral Fellowship, Chulalongkorn University and the Chulalongkorn Academic into Its 2nd Century Project Advancement Project (Thailand); the Kavli Foundation; the Nvidia Corporation; the SuperMicro Corporation; the Welch Foundation, contract C-1845; and the Weston Havens Foundation (USA).

\end{acknowledgments}

\bibliography{auto_generated}
\cleardoublepage \appendix\section{The CMS Collaboration \label{app:collab}}\begin{sloppypar}\hyphenpenalty=5000\widowpenalty=500\clubpenalty=5000\vskip\cmsinstskip
\textbf{Yerevan Physics Institute, Yerevan, Armenia}\\*[0pt]
A.~Tumasyan
\vskip\cmsinstskip
\textbf{Institut f\"{u}r Hochenergiephysik, Wien, Austria}\\*[0pt]
W.~Adam, T.~Bergauer, M.~Dragicevic, J.~Er\"{o}, A.~Escalante~Del~Valle, R.~Fr\"{u}hwirth\cmsAuthorMark{1}, M.~Jeitler\cmsAuthorMark{1}, N.~Krammer, L.~Lechner, D.~Liko, I.~Mikulec, F.M.~Pitters, N.~Rad, J.~Schieck\cmsAuthorMark{1}, R.~Sch\"{o}fbeck, M.~Spanring, S.~Templ, W.~Waltenberger, C.-E.~Wulz\cmsAuthorMark{1}, M.~Zarucki
% ...
\vskip\cmsinstskip
% ...
1:  Also at TU Wien, Wien, Austria\\
% ...
\end{sloppypar}
\end{document}